\documentclass[review]{elsarticle}
\usepackage{lineno}
\usepackage[colorlinks,
            linkcolor=blue,       
            anchorcolor=blue,  
            citecolor=blue,        
            ]{hyperref}

\usepackage[utf8]{inputenc}

\usepackage{tikz}
\usepackage{amsfonts}
\usepackage{amsmath,amssymb}
\usepackage{systeme,mathtools}
\usetikzlibrary{positioning,arrows.meta,quotes}
\usetikzlibrary{shapes,snakes}
\usetikzlibrary{bayesnet}
\tikzset{>=latex}
\usepackage{pgfplots}
\UseRawInputEncoding
\DeclareMathAlphabet{\mathscr}{OT1}{pzc}{m}{it}
\usepackage{amssymb,amsmath}
\usepackage{geometry}
\usepackage{graphicx}
\usepackage{epstopdf}
\usepackage{booktabs}
\usepackage{color}
\usepackage{colortbl}
\usepackage{tabularx}
\usepackage{amsthm}                   
\newtheoremstyle{mystyle}
{3pt}    
{3pt}    
{}       
{}       
{\bfseries} 
{.}      
{.5em}   
{}       
\theoremstyle{mystyle}             
\newtheorem{definition}{Definition}

\usepackage{multirow}
\usepackage{graphicx}
\usepackage{subfigure}
\usepackage{tikz}
\usepackage{threeparttable}
\usepackage[ruled]{algorithm2e}

\begin{document}
\begin{frontmatter}
\title{Grey system model on  time scales}
\author[label1]{Wanli Xie}
\address[label1]{Institute of Information Management and Data Communication, School of Communication, Qufu Normal University, Rizhao 276826,China}


	\begin{abstract}
	The Grey System Theory (GST) is a powerful mathematical framework employed for modeling systems with uncertain or incomplete information. This paper proposes an integration of the GST with time scales, a generalized approach that encompasses both discrete and continuous time models. The proposed model, called the Grey System Model on Time Scales (GST-T), offers a robust solution for analyzing hybrid systems where events occur on varying time domains.
\end{abstract}
\end{frontmatter}  

\section{Introduction}
In many real-world applications, systems exhibit uncertain, incomplete, or imprecise data, making it difficult to develop accurate models using traditional methods. In such cases, the Grey System Theory, proposed by Deng \cite{deng1982control} in 1982, has proven to be a powerful tool for modeling, analyzing, and forecasting systems with sparse or uncertain data. GST focuses on the use of limited data to predict future trends, especially when full information about a system is not available. A core component of GST is the Grey Model (GM), particularly the GM(1,1) model, which has become a widely used method for time series forecasting \cite{liu2017grey}.
The GM(1,1) model is a first-order differential equation that describes the evolution of a single variable within a system. It is particularly effective in addressing systems with incomplete historical data and has been successfully applied across various fields, such as economics, engineering, environmental sciences, and social sciences. However, in the equation $\frac{{d{x^{(1)}}(t)}}{{dt}} + a{x^{(1)}}(t) = b,$
the variable \(x^{(1)}(t)\) is not rigorously defined but is instead characterized within the framework of difference equations. This transition between difference and differential equations introduces additional challenges, further limiting the generalizability of the model.

To address this limitation, a more generalized approach is required. This is where time scales theory comes into play. Time scales theory provides a unifying framework for studying dynamic systems that evolve over both continuous and discrete time domains \cite{bohner2001dynamic}. A time scale is defined as a closed subset of real numbers, which may consist of discrete, continuous, or a combination of both types of points. This generalization allows for the modeling of hybrid systems that exhibit both discrete events and continuous processes in a seamless manner.
	\section{Preliminaries and problem formulation}

\subsection{The time scales calculus}
A \textit{time scale} $\mathbb{T}$ is any nonempty closed subset of the real numbers $\mathbb{R}$. Examples include $\mathbb{R}$ (real numbers), $\mathbb{Z}$ (integers), $\mathbb{N}$ (natural numbers), finite unions of closed intervals (e.g., $[0,1] \cup [2,3]$), and the Cantor set. Sets that are not closed, such as $\mathbb{Q}$ (rationals) or open intervals like $(0,1)$, are not time scales. A time scale $\mathbb{T}$ inherits the standard topology from $\mathbb{R}$.

\begin{definition}[\cite{bohner2001dynamic}]
	Let $\mathbb{T}$ be a time scale. The forward jump operator $\sigma: \mathbb{T} \to \mathbb{T}$ is defined as
	\[
	\sigma(t) := \inf \{s \in \mathbb{T} \mid s > t\}, \quad \forall t \in \mathbb{T}.
	\] The backward jump operator $\rho: \mathbb{T} \to \mathbb{T}$ is given by
	\[
	\rho(t) := \sup \{s \in \mathbb{T} \mid s < t\}, \quad \forall t \in \mathbb{T}.
	\]
\end{definition}

\begin{definition} [\cite{bohner2001dynamic}]
	The \textit{graininess function} $\mu: \mathbb{T} \to [0, \infty)$ associated with a time scale $\mathbb{T}$ is defined as
	\[
	\mu(t) := \sigma(t) - t, \quad \forall t \in \mathbb{T}.
	\]
\end{definition}
\begin{definition} [\cite{bohner2001dynamic}]
	Let $\mathbb{T}$ be a time scale. The set $\mathbb{T}^\kappa$ is defined as
	\[
	\mathbb{T}^\kappa := 
	\begin{cases} 
		\mathbb{T} \setminus \{M\}, & \text{if } M = \sup \mathbb{T} \text{ and } M \text{ is left-scattered}, \\
		\mathbb{T}, & \text{otherwise},
	\end{cases}
	\]
	where $M = \sup \mathbb{T}$ represents the supremum of the time scale $\mathbb{T}$.
\end{definition}
\begin{definition} [\cite{bohner2001dynamic}]
	Let $f: \mathbb{T} \to \mathbb{R}$ be a function on a time scale $\mathbb{T}$, and let $t \in \mathbb{T}^\kappa$, where $\mathbb{T}^\kappa$ is the set of points in $\mathbb{T}$ excluding right-scattered maximums. The \textit{delta derivative} of $f$ at $t$, denoted $f^{\Delta}(t)$, is the unique value (if it exists) such that, for any $\varepsilon > 0$, there exists a neighborhood $U$ of $t$ satisfying 
	\[
	\left| f(\sigma(t)) - f(s) - f^{\Delta}(t)(\sigma(t) - s) \right| \leq \varepsilon |\sigma(t) - s|, \quad \forall s \in U.
	\]
\end{definition}
\begin{definition} [\cite{bohner2001dynamic}]
	Let $f: \mathbb{T} \to \mathbb{R}$ be a regulated function. A function $F: \mathbb{T} \to \mathbb{R}$ is called a \textit{pre-antiderivative} of $f$ if it satisfies
	\[
	F^{\Delta}(t) = f(t), \quad \forall t \in \mathbb{T}^\kappa,
	\]
	where $F^{\Delta}(t)$ denotes the delta derivative of $F$.
\end{definition}
\begin{definition} [\cite{bohner2001dynamic}]
 Let $f: \mathbb{T} \to \mathbb{R}$ be rd-continuous. Then there exists an antiderivative $F: \mathbb{T} \to \mathbb{R}$ such that
	\[
	F(t) = \int_{t_0}^t f(\tau) \Delta \tau, \quad \forall t \in \mathbb{T}
	\]
	for any fixed $t_0 \in \mathbb{T}$.
\end{definition}
\begin{definition} [\cite{bohner2001dynamic}]
	A function \( p: \mathbb{T} \to \mathbb{R} \) is called \(\nu\)-\textit{regressive} if 
	\[
	1 - \nu(t)p(t) \neq 0, \quad \forall t \in \mathbb{T}^\kappa.
	\]
	The set of all \(\nu\)-regressive  is denoted by 
	\[
	\mathcal{R}_\nu = \{p: \mathbb{T} \to \mathbb{R} : p \text{ is ld-continuous and \(\nu\)-regressive}\}.
	\]
\end{definition}
\begin{definition} [\cite{bohner2001dynamic}]
	Let \( h > 0 \), set
	\begin{equation}
		Z_h := \left\{ z \in \mathbb{C} : -\frac{\pi}{h} < \operatorname{Im}(z) < \frac{\pi}{h} \right\}, \quad C_h := \left\{ z \in \mathbb{C} : z \neq \frac{1}{h} \right\}.
	\end{equation}
	Then the \(\nu\)-cylinder transformation \(\hat{\xi}_h : C_h \to Z_h\) is given by
	\begin{equation}
		\hat{\xi}_h(z) := -\frac{1}{h} \operatorname{Log}(1 - z h),
	\end{equation}
	where \(\operatorname{Log}\) is the principal logarithm function. For the degenerate case \( h = 0 \), the transformation is defined as \(\hat{\xi}_0(z) := z\) for all \( z \in \mathbb{C}_0 := \mathbb{C} \).
\end{definition}

\begin{definition} [\cite{bohner2001dynamic}]
	Consider \( p \in \mathcal{R}_\nu \), where \(\mathcal{R}_\nu\) denotes the set of appropriately defined functions. The nabla exponential function is then formally defined as
	\begin{equation}
		\hat{e}_p(t, s) := \exp\left( \int_s^t \hat{\xi}_{\nu(\tau)}(p(\tau)) \nabla \tau \right)
	\end{equation}
	for all \( s, t \in \mathbb{T} \).
\end{definition}
\subsection{Grey system model on time scales} 
Let \( x(t) \) be the original data function defined on \( \mathbb{T} \). The accumulated generating operation (AGO) is applied to the data as follows:
\[y(t) := y(t_0) + \int_{t_0}^{t} x(\tau) \Delta \tau,
\]
For continuous time (\( \mathbb{T} = \mathbb{R} \)), the integral becomes:
\[
y(t) = y(t_0) + \int_{t_0}^{t} x(\tau) d\tau.
\]
For discrete time (\( \mathbb{T} = \mathbb{Z} \)), the integral is replaced by a summation:
\[
y(t) = \sum_{\tau=t_0+1}^{t} x(\tau).
\]
The core equation for the GM(1,1) model on time scales is derived by replacing the regular derivative with the \(\Delta\)-derivative. The resulting equation  can be defined as 
\begin{equation}
	y^{\Delta}(t) + a y(t) = b.
	\label{eqn:GMont}
\end{equation}
The solution of equation (\ref{eqn:GMont}) can be derived by
\[
y(t) = 	\hat{e}_{-a}(t, t_0) \left[ y(t_0) - \frac{b}{a} \right] + \frac{b}{a}.
\]
For continuous time ($T = \mathbb{R}$), the solution of the model takes the form $e_{-a}(t, t_0) = e^{-a(t-t_0)}\left[ y(t_0) - \frac{b}{a} \right] + \frac{b}{a}$, which is consistent with the solution of the classical GM(1,1) model. For discrete time ($T = \mathbb{Z}$), the solution is expressed as $e_{-a}(t, t_0) = (1-a)^{t-t_0}\left[ y(t_0) - \frac{b}{a} \right] + \frac{b}{a}$.
	\section{Applications and discussion}
The classical grey model can be constructed in both continuous and discrete time domains. It is possible to establish the model exclusively in the continuous domain or solely in the discrete domain. For instance, the Riemann-Liouville (RL) integral can be used to construct a continuous  operator, and its solution can be represented in a numerical form \cite{oldham1974fractional}. The RL integral is defined as 
\[
D_{0, t}^{-r} f(t) = \frac{1}{\Gamma(r)} \int_0^t (t-s)^{r-1} f(s) \, ds, \quad r > 0,
\]
and its numerical form is given by
	\begin{equation}
		\begin{aligned}
			D_{0, t}^{-r} f\left(t_n\right) & =\frac{1}{\Gamma(r)} \sum_{k=0}^{n-1} \int_{t_k}^{t_{k+1}}\left(t_n-s\right)^{r-1} f\left(t_k\right) d s \\
			& =\frac{-1}{\Gamma(r)} \sum_{k=0}^{n-1} \int_{t_k}^{t_{k+1}}\left(t_n-s\right)^{r-1} d\left(t_n-s\right) f\left(t_k\right) \\
			& =\frac{-1}{\Gamma(r)} \sum_{k=0}^{n-1} \frac{1}{r}\left[\left(t_n-t_{k+1}\right)^r-\left(t_n-t_k\right)^r\right] f\left(t_k\right) \\
			& =\frac{-\Delta t^r}{\Gamma(r+1)} \sum_{k=0}^{n-1}-\left[(n-k)^r-(n-k-1)^r\right] f\left(t_k\right) \\
			& =\frac{\Delta t^r}{\Gamma(r+1)} \sum_{k=0}^{n-1}\left[(n-k)^r-(n-k-1)^r\right] f\left(t_k\right) \\
			& =\frac{\Delta t^r}{\Gamma(r+1)} \sum_{k=1}^n\left[(n+1-k)^r-(n+1-k-1)^r\right] f\left(t_k\right).
		\end{aligned}
	\end{equation}
This numerical form can also be expressed in matrix form as
\begin{equation}
D = \left[ {\begin{array}{*{20}{c}}
		{\Psi (1)}&{\Psi (1)}&{\Psi (1)}& \cdots &{\Psi (1)}\\
		0&{\Psi (2)}&{\Psi (2)}& \cdots &{\Psi (2)}\\
		0&0&{\Psi (3)}& \cdots &{\Psi (3)}\\
		\vdots & \vdots & \vdots & \ddots & \vdots \\
		0&0&0& \cdots &{\Psi (n)}
\end{array}} \right],
\end{equation}
where
\begin{equation}
	\Psi (k) = \frac{1}{{\Gamma (r  + 1)}}\left[ {{{(n + 1 - k)}^r } - {{(n + 1 - k - 1)}^r }} \right].
\end{equation}
\begin{figure}
	\centering
	\includegraphics[width=1 \linewidth]{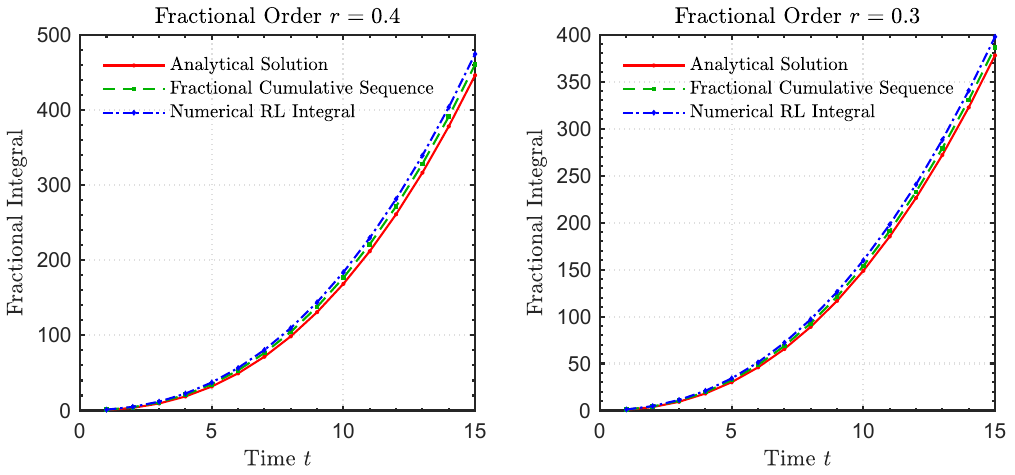}
	\caption{Fractional integral comparison for different orders.}
	\label{fig:comacc}
\end{figure}
The inverse matrix of \(D\) can be used to compute predictions based on the grey model, providing an efficient approach for obtaining forecasted values.	

In this study, we examine the fractional integral of the power function \( y(t) = t^2 \) using both analytical and numerical approaches to validate the efficacy of the proposed methods. The analytical solution is derived based on the RL integral, yielding \( I_t^r y(t) = \frac{\Gamma(3)}{\Gamma(3 + r)} t^{2 + r} \), where \( \Gamma(\cdot) \) denotes the Gamma function and \( r \) represents the fractional order. To complement the analytical results, we employ two numerical techniques: the accumulative method \cite{wu2013grey} and the Riemann-Liouville (RL) numerical integration. The accumulative method computes the fractional cumulative sum by applying weighted coefficients derived from the Gamma function, while the RL numerical integration discretizes the integral using finite difference schemes to approximate the fractional integral. The experiment is conducted over the time interval \( t \in [1, 15] \) with fractional orders \( r = 0.4 \) and \( r = 0.3 \). Comparative analysis, as illustrated in Figure~\ref{fig:comacc}, demonstrates that both numerical methods closely align with the analytical solution, confirming their accuracy and reliability. Minor discrepancies observed are attributed to inherent discretization errors in the numerical schemes.

Certainly, we can also integrate other continuous fractional calculus frameworks in conjunction with advanced numerical algorithms to systematically develop novel grey system models.

	\section{Conclusion}
	In this paper, we have introduced the grey system model on time scales , which extends the traditional  grey system model by integrating time scales theory. This extension unifies continuous and discrete time systems into a single framework, offering greater flexibility in modeling hybrid dynamic systems.




\bibliographystyle{elsarticle-num}
\bibliography{greybib}
\end{document}